\documentclass[12,twocolumn]{article}
\usepackage{amsmath}
\usepackage{listings}
\usepackage{xcolor}
\usepackage{chngcntr}
\usepackage{url}

\begin{document}
\title{Quantum pattern matching Oracle construction}

\author{Vikram Menon \\
  Ayan Chattopadhyay$^{*}$}

\twocolumn[
  \begin{@twocolumnfalse}
    \maketitle
    \begin{abstract}
  We propose a couple of oracle construction methods for quantum pattern matching. We in turn show that one of the construct can be used with the Grover's search algorithm for exact and partial pattern matching, deterministically. The other one also points to the matched indices, but primarily provides a means to generate the Hamming distance between the pattern to be searched and all the possible substrings in the input string, in a probabilistic way.
    \end{abstract}
  \end{@twocolumnfalse}
]

\subsection*{Introduction}
Pattern searching has become very crucial in current times because of its application in data analysis, be it DNA sequencing, search engines, AI, etc. These domains deal with significantly large amount of data and require pattern matching as a basic tool to extract useful information for various use cases. One of the important parameters to enhance searching performance is to efficiently pre-process the available data. There have been few attempts to exploit quantum parallelism to improve the runtime of the serach algorithms. Ramesh H. and Vinay V. initially proposed a quantum method\cite{RV} for substring matching in $\tilde{O}(\sqrt{N} + \sqrt{M})^{1}$ quantum time, combinging Grover's search algorithm\cite{GRV} and Deterministic Sampling. A quantum algorithm for closest pattern matching was given by \cite{MO}, which allows to search for as many distinct patterns in a given string using a query function per symbol of the pattern alphabet. It returns the position of the closest substring to a given pattern with non-negligible probability in $O(\sqrt{N})$ queries, where $N$ is the size of the string.

Here, we propose two oracle construction methods. We will show how the first one can be used with the Grover's search algorithm for exact pattern (substring of size $M$) matching in a given input string (size $N$). The search will return the position of one of the occurances (in case of multiple match) of the substring, in $\sqrt{N-M}$ oracle queries. We will later extend this to variable length search. The input quantum state preparation will use the naive pattern search technique. The second oracle, which doesn't require Grover's algorithm, will provide a means to generate the Hamming disance of the pattern to be searched and all the matched indices in the given text. Though it will reveal all the required information, the outcome will be probabilistic and will have similar running time as that of Grover's search. To get all the matched indices, a large number of iterations ($N-M$) need to be performed in proportion to the running time. 

\section*{Oracle construction for Grover's search algorithm}
Let us consider an input string of size $N$, over an alphabet set $\sum$. The naive search would require a pattern of size $M$ to be matched in the input string starting at the first position, and repeating it by incrementing the position by one in the forward direction, until there is a match. If '$i$' represents a search position, then $i^{th}$ substring in input string to be matched will constitute the alphabets from position $i$ to $i+M-1$, that is to say that the search positions will be in the range $i \in [0,N-M-1]$.

The input to the Grover's search will be a $t$-qubit quantum register initialized to state $|0\rangle^{\otimes t}$, where $2^{t} = T = (N-M)$. It is trivial to note that the basis state index, $0$ to $T-1$, directly maps to all the search positions in the input string. The oracle can be defined as,
\begin{equation}
  f_{M}(x) =
  \begin{cases}
    1, &         \text{if 'x' is a solution}\\
    0, &         \text{otherwise}.
  \end{cases}
\end{equation}
where, $x$ is a solution if the given pattern matches the substring of size $M$ at postion $x$ of the input string.
The oracle, as expected, will mark all the solution states as given below,
\begin{equation}
  \begin{split}
    O[H^{\otimes t}|0\rangle^{\otimes t}] &= O[\frac{1}{\sqrt{T-1}} \sum_{x=0}^{T-1} |x\rangle]\\
    &= \frac{1}{\sqrt{T-1}} \sum_{x=0}^{T-1} (-1)^{f_{M}(x)} |x\rangle.
  \end{split}
\end{equation}

The measurement will yield the solution state index that corresponds to the position of the matched pattern in the input string. In case of multiple matches, one of the indices will be returned at random. We can see that the number of oracle queries is $O(\sqrt{T}) = O(\sqrt{N-M})$, as guided by the Grover's algorithm.

The oracle above was defined to compare all the $M$ alphabets in the pattern to be searched. By modifying it to match $K$ alphabets instead of $M$, with $K < M$, a prefix pattern matching can be achieved. This may result in multiple solutions in terms of the  superposition of all the matched state indices (positions) in the final state. Running multiple iterations of the algorithm will retrieve all the solutions.

\section*{Oracle construction of other kind}
The following algorithm articulation and discussion is based on taking an arbitraty example of DNA sequening. There are four (4) types of nucleotides - Adenine(A), Thymine(T), Guanine(G), Cytosine(C), for simplicity we have omitted the Uracil, the $5^{th}$ one. Suppose
\begin{equation}
  \begin{split}
    |A\rangle &= |000\rangle\\
    |T\rangle &= |010\rangle\\
    |G\rangle &= |100\rangle\\
    |C\rangle &= |110\rangle
  \end{split}
\end{equation}
We have kept $Z \in \sum$ reserved and $|Z\rangle = |111\rangle$ for addition of any padding or junk character for mathematical function to work.
We represent the pattern, say $P$, as a tensor product of various combinations of the above four bases, for example a pattern of $AGGCA$ can be represented as $|A\rangle \otimes |G\rangle \otimes |G\rangle \otimes |C\rangle \otimes |A\rangle$.

The large input string, of size $N$, needs to be pre-processed before applying an oracle. The pre-processing will result in all possible combinations of substrings of size equal the the pattern, say $M$, in the span of input string.
Each such substring can be formed by sliding the classical text index by one. The string at the dummy / padding index would be filled with all the prohibited text ($Z$) - to be distinguishable from all the nucleotide bases in this running example.
As an example, $i^{th}$ index entry will contain the substring(say $s$) constituting the alphabets(nucleotide) $i$ to $i+M-1$ of the input string. Thus each such combination is actually formed as tensor product of $|x_{i}\rangle$, the index state with the corresponding text string state, $|s\rangle$.
  
 Multiple occurances of a substring will result in duplicate entries of string state under different index states. As an example, there can be a possibility where same $|s\rangle$ would appear in $|x_{i}\rangle \otimes |s\rangle$, and also in $|x_{j}\rangle \otimes |s\rangle$, where $i \neq j, i, j \in N$.
 
There will be at most $N-M+1$ entries and the time complexity of the pre-processing will be $O(N-M)$. Hence, the overall input string state is pre-processed and prepared as follows,
\begin{equation}
  \sum_{x=0}^{T-1} |x\rangle \otimes |s_{x}\rangle
\end{equation}
where $x \subset N$ and $s_{x}$ is the corresponding substring at index $x$ of the database. In care of multiple occurances, the same substring would appear in $|x\rangle \otimes s_{x}$ and $|y\rangle \otimes s_{y}$, i.e. $x \neq y$ and $s_{x} = s_{y}$.

The above equation can be written as follows:
\begin{equation}
  \sum_{x=0}^{T-1} |x\rangle \otimes |s_{x}\rangle = |1\rangle \otimes |s_{1}\rangle + |2\rangle \otimes |s_{2}\rangle + ... + |T-1\rangle \otimes |s_{T-1}\rangle
\end{equation}

Each $|i\rangle \otimes |s_{i}\rangle$, $i \in \{1, 2, ... T-1\}$ is found to be independent unit and can be used in parallel to fed to the oracle. This would mean that the runtime can be reduced significantly, by leveraging the parallel computation on similar set of oracle with different set of input.

The input to the oracle will include the above string state, the pattern and some working qubits having size same as pattern, i.e $M$, as given below
\begin{equation}
  |i\rangle \otimes |s_{i}\rangle, |P\rangle, |0\rangle^{\otimes m}
\end{equation}
where $2^{m} = M$ and $i \in N$.

The output of one such oracle for any of the above arrangement is defined by,
\begin{equation}
  \begin{split}
    O [|i\rangle \otimes |s_{i}\rangle, &|P\rangle, |0\rangle^{\otimes m}] = \\
    &|i\rangle \otimes |s_{i}\rangle, |P\rangle, |0\rangle^{\otimes m} \oplus |P\rangle \oplus |s_{i}\rangle
  \end{split}
\end{equation}

Now measure $3^{rd}$ register, the output register. A value of '$0$' indicates an exact match, with the $1^{st}$ register containing the states that are consistent with the output. The partial measurent on the $1^{st}$ register would produce one of the matching indices of the given pattern within the input string. To get all the matched indices all $(N-M)$ iterations will need to be measured. A non-zero value will indicate a partial match and will represent the Hamming distance between the searched pattern and the substring. The measurement result on a large ensemble of input states will provide a map of how close the substring to be searched is with all the possible substrings in the input string. 

The parallel run of oracle would greatly enhance running time. The amortized running time would be $O(M)$ which is equal to length of the pattern as the unit consisting of preprocessing and the oracle can be run in parallel. Any $|i\rangle$ state in the defined range (need not be sequential) can be fed to the unit above. By birthday paradox the average running time for the searched pattern can be found out in $O(\sqrt{N})$.

Suppose, if the above described oracle cannot be used in parallel, that means if just one unit per overall search operation is allowed, then the following approach can be thought of as an alternative. 
From Equation (5) the arrangement of $\sum_{x=0}^{T-1} |x\rangle \otimes |s_{x}\rangle$ is fed to the oracle instead. This can be achieved by adding all the pre-processed states and which is nothing but $|1\rangle \otimes |s_{1}\rangle +  |2\rangle \otimes |s_{2}\rangle + ...+ |T-1\rangle \otimes |s_{T-1}\rangle$. As $\forall i, |i\rangle$ remains the basis states, the above construction is possible. The advantage of parallel computation however is lost, but in this construction any of the matched index can be found with keeping the same running time as above. The disadvantage is to get all the indices all the $N-M$ iterations will be tried by repeating the same pattern threrby increasing running time by large proportion.

\section*{Conclusion}
We have shown two methods of constructing oracles that can be usefull for pattern searching. In the first case, the oracle when used in the Grover's search algorithm provided a means to perform exact and partial pattern matching. In addition, the search space dimension was reduced from $N$ to $N-M$, thereby reducing the number of oracle queries. There could be other unique ways of preparing oracle and the input state, which may further improve the outcome and runtime when combined with the Grover's algorithm. 

The second oracle by its construction gives the opportunity of parallel processing in the independent similar oracle units. This oracle also produces all the matched indices of the pattern from the large input text although the running time would be higher than the first method. This construction can be used in providing a probabilistic pattern search method to keep similar running time as of Grover and a means to calculate the Hamming distances of all the the searched substring with the searched pattern. Measurements on a large ensemble of input states will provide a map or signature of how close the pattern to be searched is with respect to all the possible substrings in the input string.
Thus it is observed that second oracle construction can have wide level of use in pattern searching.  

%% The authors would like to thank Dr. Rajendra K. Bera for useful discussions and inputs.


\begin{thebibliography}{1}

\bibitem{RV}
H. Ramesh and V. Vinay. String matching in O(n+m) quantum time. J. Discrete Algorithms {\bf 1}, (2003) 103-110.

\bibitem{GRV}
Grover. L, A fast quantum mechanical algorithm for database search. Proceedings of 28th ACM Symposium on Theory of Computing, 1996, pp. 212-219.  
\bibitem{MO}
Mateus. P. and Omar. Y., Quantum Pattern Matching, 2005. arXiv:quant-ph/0508237.

%% \bibitem{CA}
%%  G. Navarro, ACM Comput. Surv. {\bf 33}, 31 (2001), ISSN0360-0300.
\end{thebibliography}
\end{document}